\documentclass[twocolumn]{aastex631}

\usepackage{multirow}
\usepackage[dvipsnames]{xcolor}

\definecolor{blue}{HTML}{1f77b4}
\definecolor{orange}{HTML}{ff7f0e}
\definecolor{green}{HTML}{2ca02c}
\definecolor{red}{HTML}{d62728}

\definecolor{pink}{HTML}{e377c2}
\definecolor{brown}{HTML}{8c564b}
\definecolor{olive}{HTML}{bcbd22}
\definecolor{purple}{HTML}{9467bd}

\shorttitle{CO second overtone variability in three post-AGB stars}
\shortauthors{Puķītis}

\graphicspath{{./}}

\begin{document}

\title{CO second overtone line variability in three carbon-rich early post-AGB stage stars}

\correspondingauthor{Kārlis Puķītis}
\email{karlis.pukitis@lu.lv}

\author[0000-0003-2599-6126]{Kārlis Puķītis}
\affiliation{Laser Center, Faculty of Science and Technology,
University of Latvia, Raiņa bulvāris 19, LV-1586 Rīga, Latvia}

\begin{abstract}

Near-infrared CO molecular lines are suggested to be linked with dynamic processes in post-AGB stars; however, they have not been investigated at high spectral resolution. CO second overtone line variability in the H-band is presented for the cool carbon-rich post-AGB stars IRAS 22272+5435, IRAS Z02229+6208, and IRAS 20000+3239. CO features are observed to change intensity, shape, and radial velocity as well as switch between emission and absorption during the pulsation cycle. At all times, the CO line positions are located no more than around 10 km/s away from the systemic velocity. Molecular line variation is qualitatively similar in all three stars, and it appears that the site of formation is the extended atmosphere. Emission in CO lines tends to be strongest during pulsation phases close to light maximum, and weakest emission or absorption tends to be seen when close to light minimum, resembling the behaviour in CN Red and C\textsubscript{2} Swan system lines at shorter wavelengths. The connection of molecular line variability with pulsation of the star could be related to a shock that traverses the atmosphere once per pulsation cycle.

\end{abstract}


\section{Introduction} 
\label{sec:intro}

The post-asymptotic giant branch (post-AGB) stage is a brief phase of stellar evolution for low- and intermediate-mass stars (1-8 $M_\odot$) immediately succeeding AGB stars and preceding the planetary nebula (PN) stage. Stellar wind in the post-AGB stage contributes to shaping of PN both directly and indirectly. As this outflow is faster, it catches up and interacts with the previously ejected slower AGB wind that forms the bulk of the PN. In the so-called early post-AGB phase, evolutionary rate of a star is highly dependent on mass loss rate \citep{Bertolami2016}. If temperature of the star increases too slowly, the ejected matter will disperse before it is ionized. On the other hand, evolution that is too quick causes the ionization to occur within the surrounding nebula that is still dense and does not allow for optical radiation to escape. The knowledge of wind properties in the post-AGB stage is very limited. Spectroscopy provides a way to observe processes that are related to formation of the stellar wind in the extended atmospheres of these pulsating supergiants.

One of the features of interest for investigation of dynamic processes in post-AGB objects are the CO molecule first overtone ($\Delta v=2$) lines in the K-band. In these wavelengths, multiple post-AGB objects were investigated by \citet{Hrivnak1994}. CO features were detected for a part of them and seen either in absorption or emission. In the case of IRAS 22272+5435, a change from emission to absorption was observed, and interpreted to be caused by an episode sporadic of mass ejection with emission originating either close to the star or farther in the circumstellar envelope. The variability in the $\Delta v=2$ bands for this star was observed again and discovered for few other post-AGB objects by \citet{Raman2008}. CO first overtone bands were observed for several post-AGB stars also by \citet{Oudmaijer1995}. In the case of HD 170756 (AC Herculis), a transformation from emission to absorption was detected, and attributed to expanding and infalling layers of the outer atmosphere. Investigation of post-AGB objects in the near-infrared being mostly limited to low- or moderate spectral resolution prohibits shaping firm conclusions about the site of CO line formation. \citet{Mozurkevizh1987} followed the variability of the K-band CO lines in the RV Tauri type variable R Scuti. They used R=$\Delta\lambda$/$\lambda\approx$30000 spectra that allowed to resolve individual CO lines. The complex, variable profiles showing both emission and absorption were concluded to form in the upper atmosphere.

The near-infrared region houses also the second overtone ($\Delta v=3$) lines of the CO molecule. Specifically, these are located in the H-band. Despite this wavelength region having been investigated for many post-AGB objects \citep{Hrivnak1994,Oudmaijer1995,Raman2008}, detection of CO $\Delta v=3$ lines has been reported only for IRAS 22272+5435 by \citet{Raman2008}. While CO $\Delta v=2$ features were observed in absorption and after one and a half months they were absent, no pronounced changes were seen in the relatively weak $\Delta v=3$ absorptions. \citet{Raman2008} reported weak CO second overtone absorption also for V450 Vul; however, it is likely that this star is not a post-AGB object. In this study, CO $\Delta v=3$ lines and their variability is explored in high-resolution for three post-AGB stars with the aim to clarify the site of formation for CO near-infrared features.

\begin{figure}
\includegraphics[width=\linewidth, trim={0.2cm 0.7cm 1.5cm 1.0cm},clip]{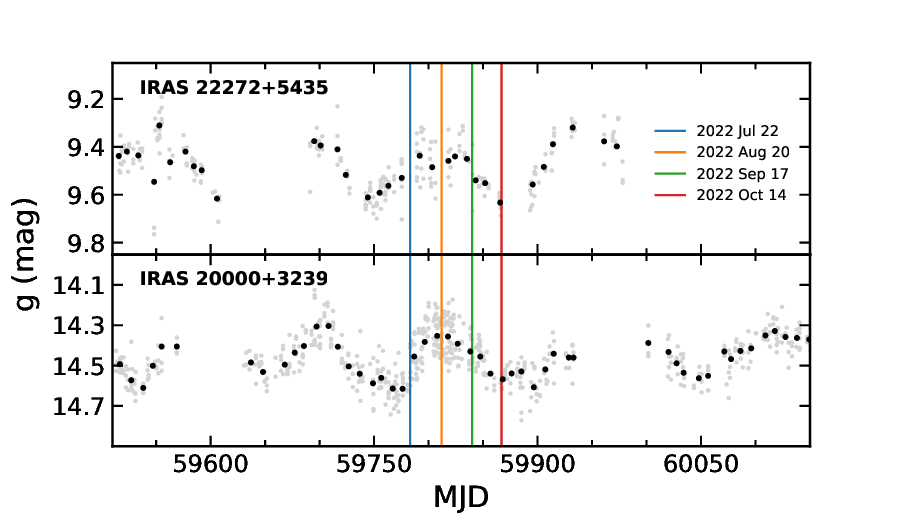}
\includegraphics[width=\linewidth, trim={0.3cm 0.9cm 0.35cm 0.35cm},clip]{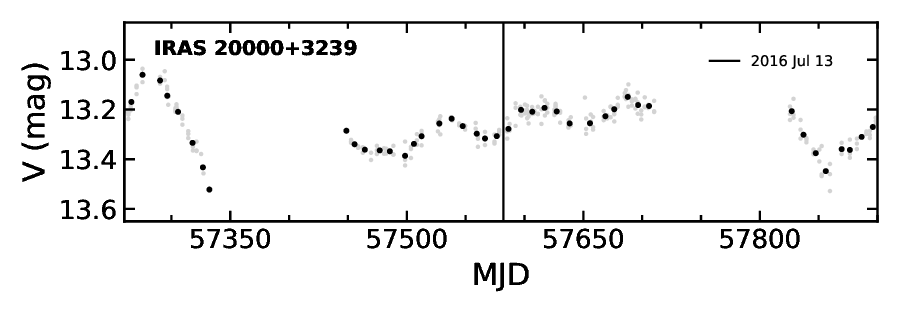}
\includegraphics[width=\linewidth, trim={0.3cm 0.9cm 0.35cm 0.35cm},clip]{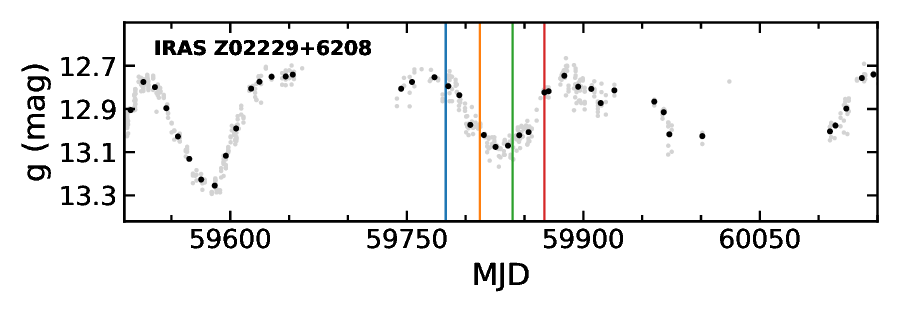}
\includegraphics[width=\linewidth, trim={0.3cm 0.35cm 0.45cm 0.35cm},clip]{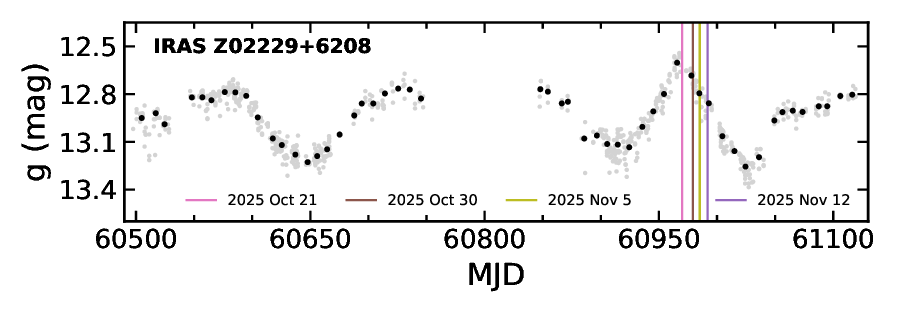}
\caption{ASAS-SN light curves of the three post-AGB stars around the time of the spectroscopic observations. Gray points show all photometric measurements with outliers and lower precision data (error $>$ 0.05$^m$) removed, and black points show 10-day binned data. Only "bt" camera data are used in the case of IRAS 22272+5435. Relatively high scatter in the light curve for this object is due to its higher brightness leading to saturation effects.}
\label{fig:lightcurves}
\end{figure}

\begin{deluxetable*}{ccccccc}
\tablecaption{CARMENES observation dates, exposure times, corresponding photospheric radial velocities in heliocentric scale and with respect to the systemic velocity, and pulsation phases\label{tab:rv}}
\tablehead{
\colhead{} & \colhead{Date} & \colhead{MJD} & \colhead{$t_{exp}$} & \colhead{${V_{r, helio.}}$} & \colhead{${\Delta V_{r, syst.}}$}& \colhead{$\varphi$} \\
\nocolhead{} & \nocolhead{} & \nocolhead{} & \nocolhead{} & \colhead{(km/s)} & \colhead{(km/s)} & \nocolhead{}}
\startdata
IRAS 22272+543    & \textcolor{blue}{2022 Jul 22} & 59783.1 & 3 min             & -34.8  &  6.0 & 0.29\\
                  & \textcolor{orange}{2022 Aug 20} & 59812.0 & 3 min             & -38.1  &  2.7 & 0.50\\
                  & \textcolor{green}{2022 Sep 17} & 59840.0 & 3 min             & -42.2  & -1.4 & 0.71\\
                  & \textcolor{red}{2022 Oct 14} & 59866.9 & 3 min             & -41.4  & -0.6 & 0.91\\
IRAS Z02229+6208  & \textcolor{blue}{2022 Jul 22} & 59783.1 & 5 min             &  23.8  &  3.7 & 0.60\\
                  & \textcolor{orange}{2022 Aug 20} & 59812.2 & 5 min             &  18.7  & -1.4 & 0.84\\
                  & \textcolor{green}{2022 Sep 17} & 59840.0 & 5 min             &  24.4  &  4.3 & 0.07\\
                  & \textcolor{red}{2022 Oct 14} & 59866.9 & 5 min             &  25.3  &  5.2 & 0.30\\
                  & \textcolor{pink}{2025 Oct 21} & 60969.9 & 7 min             &  17.7  & -2.4 & 0.53\\
                  & \textcolor{brown}{2025 Oct 30} & 60979.1 & 3 $\times$ 7 min  &  16.7  & -3.4 & 0.61\\
                  & \textcolor{olive}{2025 Nov 5} & 60985.1 & 4 $\times$ 11 min  &  15.5  & -4.6 & 0.66\\
                  & \textcolor{purple}{2025 Nov 12} & 60992.0 & 4 $\times$ 11 min &  15.4  & -4.7 & 0.72\\
IRAS 20000+3239   & 2016 Jul 13 & 57582.1 & 30 min            &  -3.9  &  1.2 &    -\\
                  & \textcolor{blue}{2022 Jul 22} & 59783.1 & 15 min            &  -0.4  &  4.7 & 0.24\\
                  & \textcolor{orange}{2022 Aug 20} & 59812.0 & 15 min            &  -3.5  &  1.6 & 0.50\\
                  & \textcolor{green}{2022 Sep 17} & 59840.0 & 15 min            &  -2.8  &  2.3 & 0.75\\
                  & \textcolor{red}{2022 Oct 14} & 59866.9 & 15 min            &  -3.7  &  1.4 & 0.00
\enddata
\end{deluxetable*}

\section{Observations} 
\label{sec:obs}

The stars IRAS 22272+5435, IRAS Z02229+6208, and IRAS 20000+3239 were chosen for investigation of CO second overtone line variability. For the first one, the $\Delta v=3$ lines have already been detected and the $\Delta v=2$ lines are known to be variable. Absorption in CO first overtone lines has been detected also for the other two objects \citep{Hrivnak1994,Raman2008}. The stars are very similar - they are carbon and s-processed enriched G-type supergiants \citep{DeSmedt2016,Reddy1999,Klochkova2006} in the early stage of post-AGB evolution as indicated by their relatively low surface temperatures (5750 K or cooler; \citet{Zacs2025, Reddy1999, Klochkova2006}) and long pulsation periods (around 130 days or longer; \citet{Hrivnak2022}). 

Spectra of the three stars were observed with the CARMENES instrument \citep{Quirrenbach2014, Caballero2025} at the 3.5m telescope at the Calar Alto observatory. CARMENES consists of VIS and NIR spectrographs with spectral coverages from 0.52 to 0.96 $\mu$m and from 0.96 to 1.71 $\mu$m. While the coverage for VIS spectrograph is continuous, in the case of NIR there is a narrow gap in every spectral order and wider gaps between the orders starting at around 1.14 $\mu$m. For the VIS spectrograph R=94.600, and R=80.400 for the NIR spectrograph. The primary observations were carried out on 2022 \textcolor{blue}{July 22}, \textcolor{orange}{August 20}, \textcolor{green}{September 17}, and \textcolor{red}{October 14}. Additional observations of IRAS Z02229+6209 were carried out on 2025 \textcolor{pink}{October 21}, \textcolor{brown}{October 30}, \textcolor{olive}{November 5}, and \textcolor{purple}{November 12}. Also, standard stars were observed in order to remove telluric absorption lines from the spectra of post-AGB objects. On the last three dates, more than one spectrum of IRAS Z02229+6208 were observed in order to look for intranight spectroscopic variability which is not detected. In the VIS range, these spectra are summed to increase signal-to-noise ratio (S/N). In the near-infrared, also individual spectra are used as they in some cases allow for a better removal of telluric absorptions. The observed spectra were reduced automatically with the CARCAL pipeline \citep{Caballero2016}. An additional CARMENES spectrum of IRAS 20000+3239 together with spectrum of a telluric standard star was downloaded from the Calar Alto Archive. These two spectra were observed on 2016 July 13. The spectra are analysed by using the software package DECH \citep{Galazutdinov2022}. To reduce noise, all spectra are smoothed with a Gaussian with full width at half maximum (FWHM) corresponding to R=50000. The metal line absorptions and low S/N in visible wavelengths and numerous molecular features make continuum placement challenging. The temporal variability of spectral features helps to select continuum points by choosing places in spectra that appear to show no variability in addition to not being affected by any constant absorption or emission. The chosen continuum points are fitted by a spline, and the spectra are normalized by dividing them with the spline.

The All-Sky Automated Survey for Supernovae (ASAS-SN; \citet{Shappee2014,Kochanek2017}) light curves (Figure \ref{fig:lightcurves}) are used to assign pulsation phases to the observed spectra. This is done by using light minimum ($\varphi$=0) or maximum ($\varphi$=0.5) dates closest to the observation times and pulsation periods. For the latter, 135 days is roughly assumed in the case of IRAS 22272+5435, 120 days - for IRAS Z02229+6208, and 110 days - for IRAS 20000+3239. The periods are estimated by following the pulsation behaviour in pulsation cycles when observations were carried out and in the neighbouring cycles. While there is an agreement with the previous pulsation period estimate for IRAS 22272+5435, \citet{Hrivnak2022} derived longer periods for the other two stars. Most likely, the disagreement between the periods is simply a consequence of semiregular pulsation nature of these stars. In the case of IRAS 20000+3239 archival spectrum, the nature of the star's pulsation at that time does not allow to assign a pulsation phase.

\begin{figure*}[htbp]
\centering
\includegraphics[width=\linewidth, trim={2.9cm 0.3cm 3cm 0cm},clip]{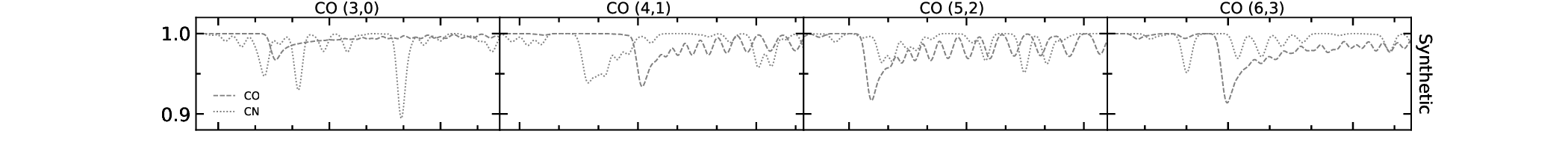}
\includegraphics[width=\linewidth, trim={2.9cm 4.9cm 3cm 3.64cm},clip]{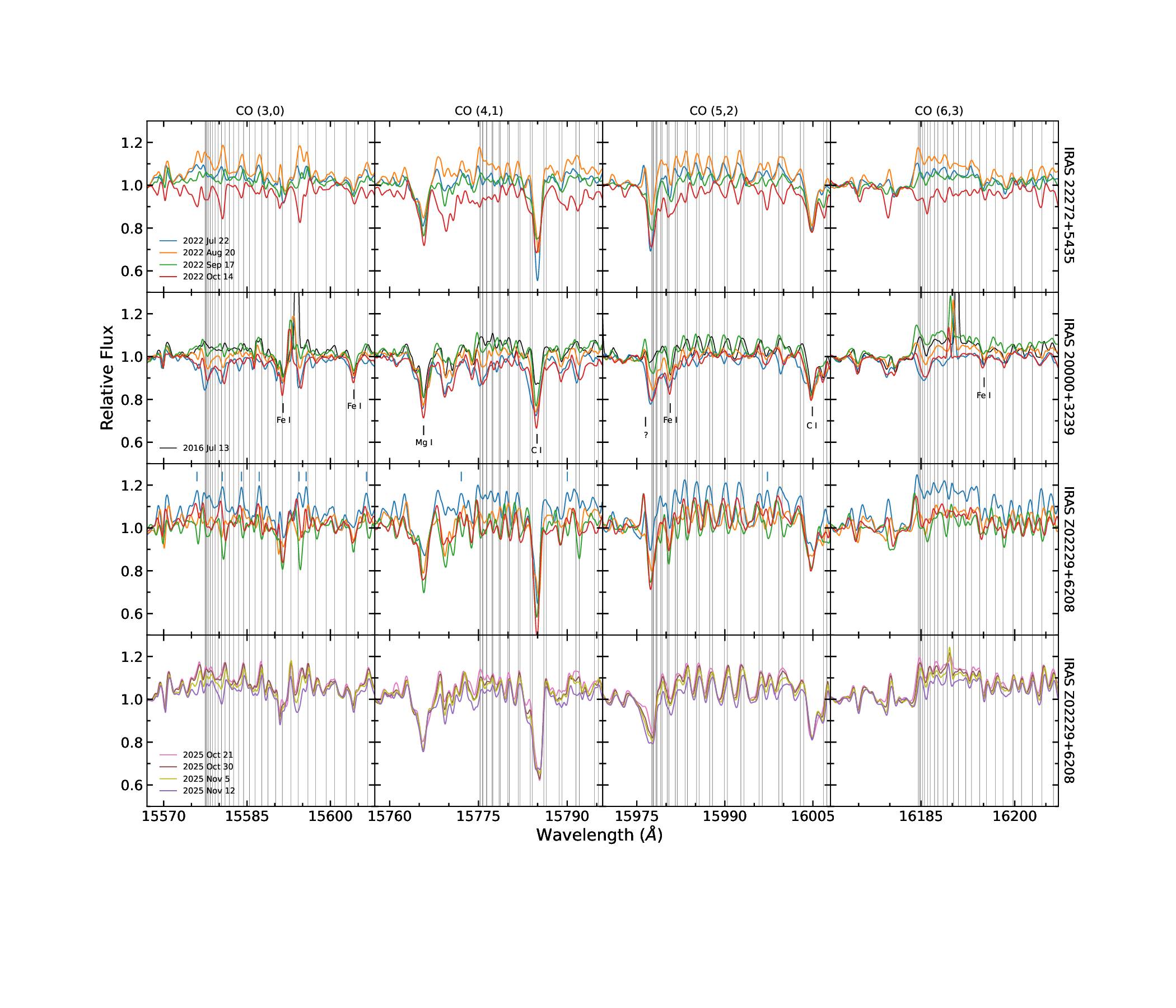}
\caption{Spectroscopic variability in the CO second overtone (3,0), (4,1), (5,2), and (6,3) bandheads for IRAS 22272+5435, IRAS Z02229+6208, and IRAS 20000+3239. Different color spectra correspond to different observation dates. Positions of CO lines belonging to the particular vibrational bandhead are marked with gray vertical lines. In the top panel, CO and CN molecule synthetic spectrum is shown for the light minimum phase of IRAS 22272+5435. Above IRAS Z02229+6208 spectra from 2022, some lines that are not a part of the CO bandhead are marked with vertical blue lines. These are either CN Red system lines or CO lines that belong to a different vibrational band. Below IRAS 20000+3239 spectra, photopsheric atomic absorptions as well as few unidentified features with unusual variability are marked. Telluric absorption lines are removed. The very intense emissions in IRAS 20000+3239 spectra are of telluric origin. To a lesser degree, these affect also IRAS Z02229+6208 spectra from 2025.}
\label{fig:CObandfirst}
\centering
\end{figure*} 

In the visible light part of the spectra, a large number of s-process element lines are seen for all three stars. Therefore, for the measurement of photopsheric radial velocity, lines that are relatively unaffected by blending are used. These lines were selected by \cite{Zacs2025} for investigation of photospheric parameter variation in IRAS 22272+5435. The measurement of radial velocity was done by cross-correlating individual line profiles with their mirror profiles. No less than 30 lines were measured in every spectrum. The estimated error for the average radial velocity is no larger than 1 km/s. Photospheric radial velocities in heliocentric scale and with respect to the systemic (stellar center of mass) velocities, as well as the assigned pulsation phases are shown in Table \ref{tab:rv}. In the case of IRAS 222272+5435, for the systemic velocity a value of -40.8 km/s is used, and it is the average photospheric radial velocity measured over the time span of 23 years \citep{Hrivnak2013}. For IRAS Z02229+6208 and IRAS 20000+3239 the adopted values of 20.1 km/s \citep{Hrivnak1999,Hrivnak2005} and -5.1 km/s \citep{Contreras2012} are based on CO radio line observations. \citet{Zacs2023} analysed the same spectra of IRAS Z02229+6208 from 2022. The result for radial velocities was slightly different than in this study (by around 0.5 km/s) due to different choice of measured lines.

\begin{figure*}[htbp]
\centering
\includegraphics[width=\linewidth, trim={2.9cm 0.3cm 3cm 0cm},clip]{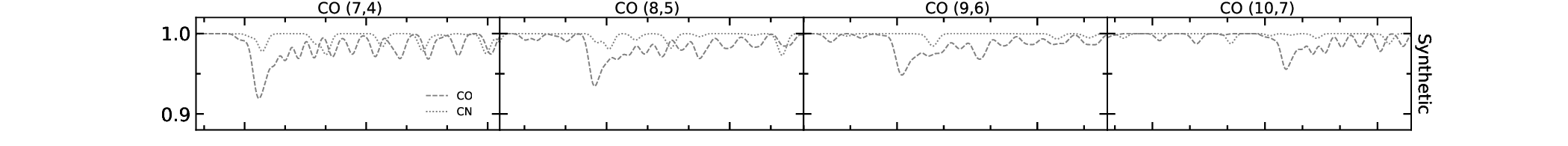}
\includegraphics[width=\linewidth, trim={2.9cm 4.9cm 3cm 3.64cm},clip]{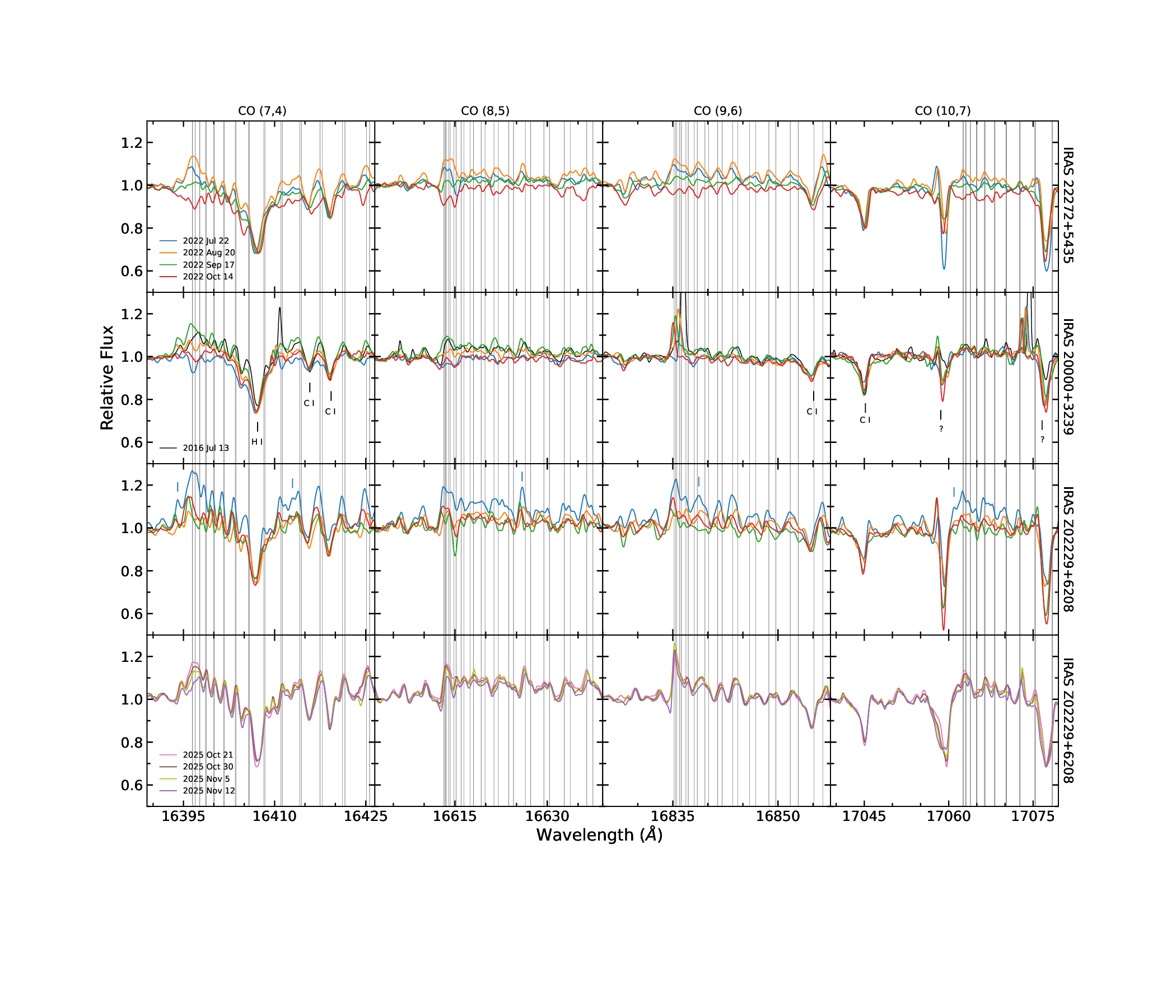}
\caption{Same as Figure \ref{fig:CObandfirst}, but for CO (7,4), (8,5), (9,6), and (10,7) bands.}
\centering
\label{fig:CObandsecond}
\end{figure*} 

\section{Analysis}
\label{sec:Analysis}

\subsection{Variation in CO bandheads and individual lines}
\label{sec:COlines}

CO features in the observed spectra of post-AGB stars are seen longward of 15577 \AA\footnote{Wavelength in air.}. All the second overtone bandheads up to CO (10,7)\footnote{The observed spectra do not reach higher vibrational bands.} are visible in spectra of the observed stars and are shown in Figures \ref{fig:CObandfirst} and \ref{fig:CObandsecond}. An exception is the CO (10,7) bandhead which is not visible in IRAS 20000+3239 spectra. For identification of CO lines, the line list by \citet{Li2015} is used.

\begin{figure}[htbp]
\centering
\includegraphics[width=\linewidth, trim={0.4cm 1.6cm 1.2cm 1.9cm},clip]{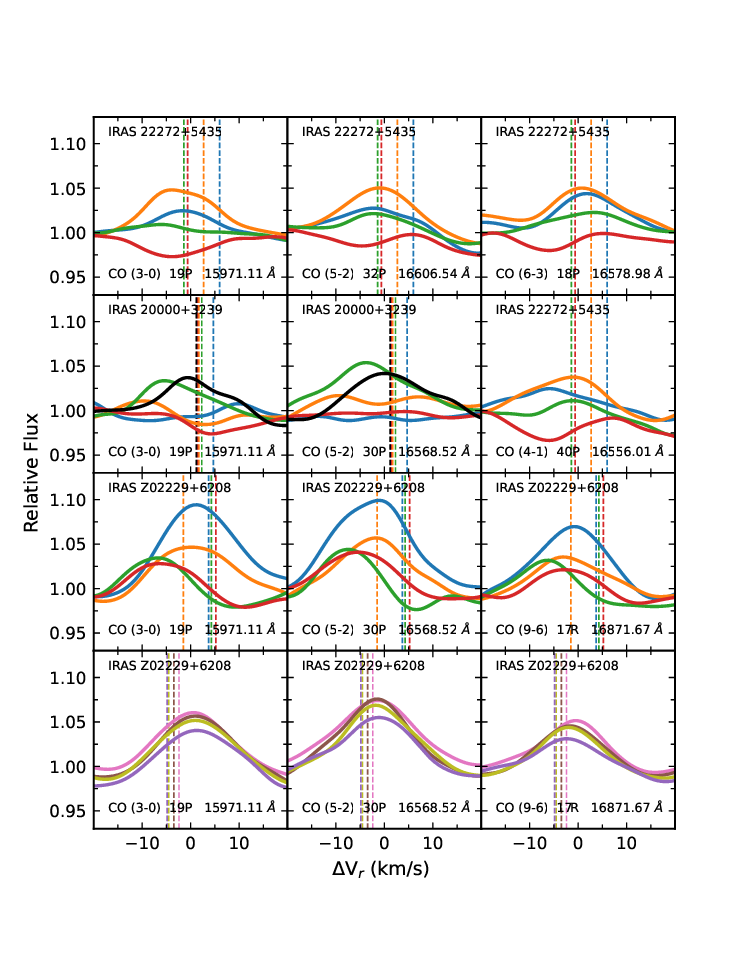}
\caption{Variability of individual CO $\Delta v=3$ lines in radial velocity scale relative to systemic velocity. Along with the vibrational band and the rest wavelength, also the rotational quantum number of the lower level ($J''$) and the branch of the transition is shown. Smoothing of the spectra does not significantly affect shape of the lines.}
\label{fig:COindivid}
\centering
\end{figure} 

The intensities of the band heads are variable. For IRAS 22272+5435, the bandheads are seen in absorption on 2022 \textcolor{red}{October 14} ($\varphi$=0.91) and in emission when the star is in pulsation phases closer to the maximum light. The emission is most pronounced on \textcolor{orange}{August 20} ($\varphi$=0.50). In the case of IRAS Z02229+6208, the emission is seen on every date and it is especially pronounced on 2022 \textcolor{blue}{July 22} ($\varphi$=0.60). On \textcolor{green}{September 17} ($\varphi$=0.07), the emission appears to be least intense and even some weak absorptions are seen. For IRAS 20000+3239, most intense absorption in the band heads is seen on 2022 \textcolor{blue}{July 22} ($\varphi$=0.24) and slightly weaker - on \textcolor{red}{October 14} ($\varphi$=0.00). The emission is most intense on \textcolor{green}{September 17} ($\varphi$=0.75). 
 
To a higher or lesser degree, every CO $\Delta v=3$ band head suffers from blending. To identify the blending features, molecular (Kurucz; based on the observed energy levels) and atomic line lists from the Vienna Atomic Line Database (VALD) database \citep{ Piskunov1995, Kupka1999} are used. Additionally, synthetic CO and CN molecule spectra are computed with the code SPECTRUM \citep{Gray1994} for the light minimum phase of IRAS 22272+5435 (effective temperature of \mbox{5050 K}; see \citet{Zacs2016} and \citet{Zacs2025} for details) to better assess the contribution of CN lines in the CO band heads. 
The CO (3,0) band head, starting at around 15577 \AA\vspace{0pt}, is not very pronounced in the surrounding spectral region due to CN Red system $\Delta v=-1$ lines. These do not only blend the CO lines in the band head but are also intense just before it, forming an emission/absorption at 15576 \AA\vspace{0pt} as strong as the band head. There are also two photospheric Fe I absorption lines in the band head. 
Right before the CO (4,1) band head, starting at around 15775 \AA\vspace{0pt}, there also are CN Red features. These are especially intense after the Mg I line at around 15766 \AA\vspace{0pt}. The band head is blended by C I absorption line.
Apart from Fe I and C I absorption lines as well as CN features, the CO (5,2) band head is affected by unidentified variable absorption that at some epochs is of inverse P Cygni type shape. The position of the absorption coincides with the expected beginning of the band head at around 15977.5 \AA\vspace{0pt}, and the emission component is located 1.0-1.4 \AA\vspace{0pt} shortward.
The start of the CO (6,3) band head is clearly seen at around at around 16184 \AA\vspace{0pt}. This band head is affected by a weak Fe I absorption and, according to synthetic spectrum, a few CN features.
The CO (7,4) band head starts at around 16396 \AA\vspace{0pt}. Approximately 2 \AA\vspace{0pt} shortward there is CO (4,1) line. The band head is mainly affected by hydrogen Br12 line at 16407 \AA\vspace{0pt}.
The CO (8,5) and (9,6) band heads start at around 16613 and 16835 \AA\vspace{0pt}, respectively. These are relatively free from blending. There seems to be an unidentified weak absorption at 16615 \AA\vspace{0pt}, well visible in 2022 \textcolor{green}{September 17} spectrum of IRAS Z02229+6208. 
Finally, the CO (10,7) band head starts at approximately 17062 \AA\vspace{0pt}. Around 1 \AA\vspace{0pt} shortward there is CO (6,3) line. Right before that, there is an unidentified line that is similar to the one in the CO (5,2) band head. There is another unidentified line at 17077 \AA\vspace{0pt} that blends the CO (10,7) band head.

It is possible to find only a few individual CO lines that appear to be not significantly blended and in a region where continuum level can be found reliably. Such lines are shown in Figure \ref{fig:COindivid}. FWHM for the CO emission lines in IRAS 22272+5435 spectra is approximately 16-17 km/s on \textcolor{blue}{July 22} and \textcolor{orange}{August 20}. On \textcolor{green}{September 17}, the width has decreased to around 14 km/s, and absorption on \textcolor{red}{October 14} shows the same width. The absorption appears to be blueshifted with respect to the systemic velocity by 5 km/s, and the emission peaks are are blueshifted by 1-2 km/s, on average. 

For IRAS Z02229+6208 both on 2022 \textcolor{blue}{July 22} and \textcolor{orange}{August 20}, the emission lines have FWHM of around 16 km/s and are blueshifted up to by around 2 km/s, on average. On 2022 \textcolor{green}{September 17} and \textcolor{red}{October 14} the emission profiles are slightly narrower as they appear to have a weak absorption on the long wavelength side resembling an inverse P Cygni-type profile. The peak of the emission component is blueshifted by roughly 6 km/s and the absorption appears to be shifted towards longer wavelengths by 10 km/s. In 2025 spectra of IRAS Z02229+6208, the emission lines have FWHM of 15 km/s and are blueshifted by up to 1 km/s, on average. These 2025 emissions ($\varphi$=0.53-0.72) are similar to the ones seen on 2022  \textcolor{orange}{August 20} ($\varphi$=0.84).

In the spectrum of IRAS 20000+3239, S/N is significantly lower than for the other two stars; therefore, FWHMs and radial velocities only for the more intense emissions from 2016 July 13 and 2022 \textcolor{green}{September 17} are measured. For the first date, the emissions are centred on the systemic velocity and for the second date they are blueshifted by around 5 km/s. Width of these emissions is around 15 km/s, on average. 

\subsection{Comparison with CN and C\textsubscript{2} spectral features}

\begin{figure*}[htbp]
\centering

\includegraphics[width=\linewidth, trim={2.9cm 0.3cm 3cm 0cm},clip]{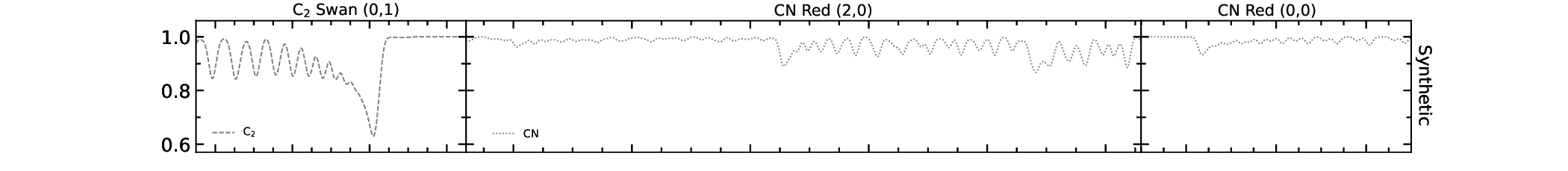}
\includegraphics[width=\linewidth, trim={2.9cm 4.8cm 3cm 3.64cm},clip]{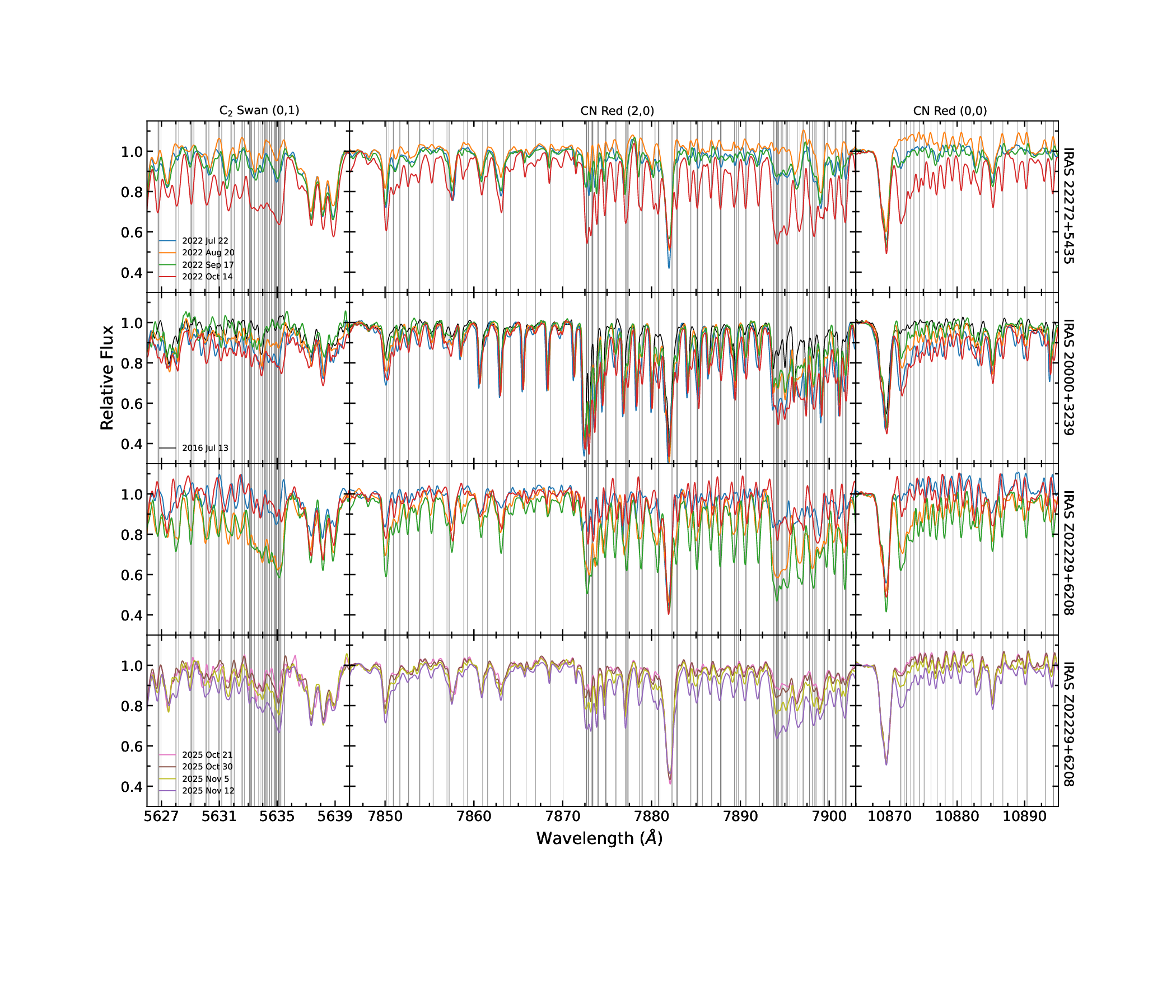}
\caption{Same as Figure \ref{fig:CObandfirst}, but for C\textsubscript{2} Swan system (1,0) and CN Red system (2,0) and (0,0) band heads. Strong Y II absorption line is visible at at 7881.9 \AA\vspace{0pt}. Strong Si I absorption is seen at 10869.5 \AA\vspace{0pt}.}
\label{fig:C2CNbandhead}
\centering
\end{figure*} 

\begin{figure}[htbp]
\centering
\includegraphics[width=\linewidth, trim={0.4cm 1.6cm 0.9cm 1.6cm},clip]{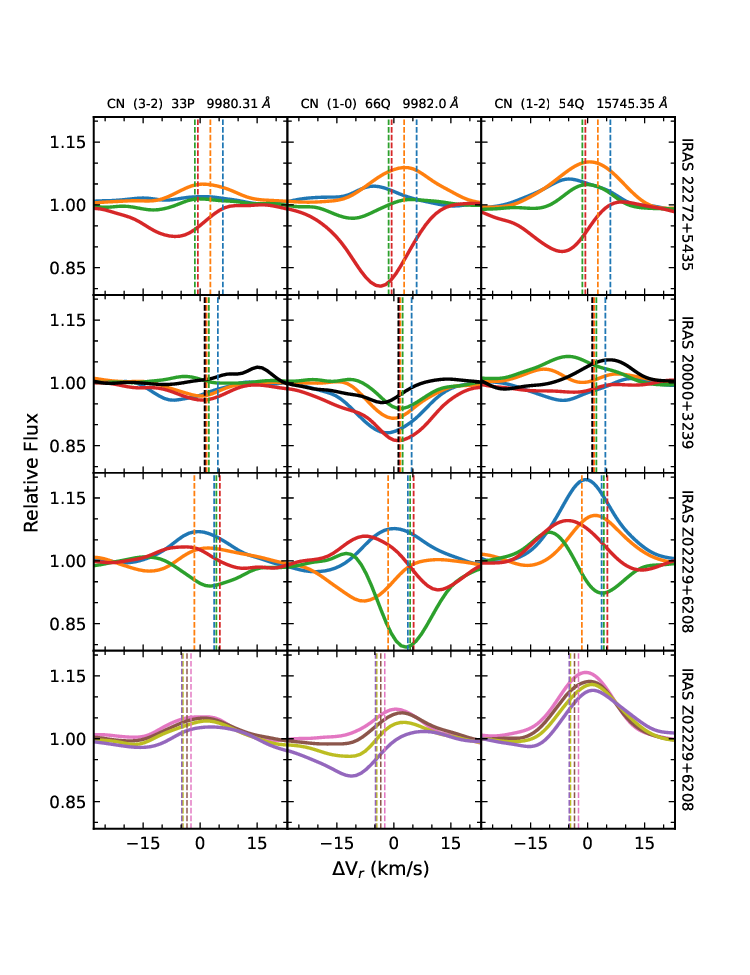}
\caption{Same as Figure \ref{fig:COindivid}, but for CN Red system lines.}
\label{fig:CNindivid}
\centering
\end{figure} 

In shorter wavelengths, band heads of other molecules are visible. Three examples are shown in Figure \ref{fig:C2CNbandhead}. For IRAS 22272+5435, the C\textsubscript{2} Swan system (0,1) and CN Red system (2,0) and (0,0) molecular line variability is similar as in the case of CO lines. On 2022 \textcolor{orange}{August 20} ($\varphi$=0.50) emission is seen and on \textcolor{red}{October 14} ($\varphi$=0.91) - absorption. On the other two dates, the molecular features are in weak emission or absorption while an emission is seen in CO lines. 

In 2022 spectra of IRAS Z02229+6208, the strongest C\textsubscript{2} and CN emissions with a similar intensity are clearly seen on 2022 \textcolor{blue}{July 22} ($\varphi$=0.60) and \textcolor{red}{October 14} ($\varphi$=0.30) while CO features only on \textcolor{blue}{July 22} are in significantly more pronounced emission than in other dates. C\textsubscript{2} and CN lines are in absorption on \textcolor{orange}{August 20} ($\varphi$=0.84) and \textcolor{green}{September 17} ($\varphi$=0.07) while CO features on these dates are mainly in emission. CO lines are seen only in emission in every 2025 spectrum but is not the case for the C\textsubscript{2} and CN lines - they are in absorption partially on \textcolor{olive}{November 5} ($\varphi$=0.66) and fully on \textcolor{purple}{November 12} ($\varphi$=0.72).

For IRAS 20000+3239, weak emissions are seen in C\textsubscript{2} Swan (0,1) and CN Red (0,0) lines on 2016 July 13 and 2022 \textcolor{green}{September 17} ($\varphi$=0.75). Strongest absorption is observed on 2022 \textcolor{blue}{July 22} ($\varphi$=0.24) and \textcolor{red}{October 14} ($\varphi$=0.0). Both these emissions and absorptions match the behaviour in CO lines.

In the case of CN Red system lines, it is also possible to find some individual lines that are not significantly affected by blending (Figure \ref{fig:CNindivid}). In IRAS 22272+5435 spectra, pure emissions are seen on 2022 \textcolor{blue}{July 22} and \textcolor{orange}{August 20}. FWHM for these is around 18 km/s, slightly higher than for CO lines. On \textcolor{blue}{July 22}, positions of CN lines are by around 3 km/s more blueshifted than CO lines, and on \textcolor{orange}{August 20} - shifted by roughly 3 km/s towards longer wavelengths. While on \textcolor{green}{September 17} CO lines are seen in emission, CN lines appear to have P Cygni-like profiles with absorption component being blueshifted by around 12 km/s with respect to the systemic velocity. The emission component is close to the systemic velocity. The CN absorption lines on \textcolor{red}{October 14} are have the same velocity as CO lines; however, the FWHH is larger - around 18 km/s. 

In IRAS Z02229+6208 spectrum from 2022 \textcolor{blue}{July 22}, CN lines have the same position and width as CO lines. While CO lines on \textcolor{orange}{August 20} are in emission, CN lines have P Cygni-like profiles with absorption blueshifted by around 12 km/s and emission peak being close to the systemic velocity. On \textcolor{green}{September 17} and \textcolor{red}{October 14}, inverse P Cygni-like behaviour is seen in both CO and CN lines; however, absorption component for the latter is more pronounced. While on \textcolor{red}{October 14} position of CN and CO lines (absorption and emission components) are similar, on \textcolor{green}{September 17} CN profiles appear to be blueshifted by around 7 km/s with respect to CO profiles. The four spectra from 2025 ($\varphi$=0.53-0.72) are very similar to and well depict the transition between 2022 \textcolor{blue}{July 22} ($\varphi$=0.60) and \textcolor{orange}{August 20} ($\varphi$=0.84) spectra. 

For IRAS 20000+3239, only the CN 15745.35 \AA\vspace{0pt} is measured since S/N is higher and emission stronger than for the other two lines. The difference when compared to CO lines seems to be larger width (around 20 km/s) on 2022 \textcolor{green}{September 17} and shift towards longer wavelengths (by around 5 km/s) in 2016 spectrum. The CN 9982.0 \AA\vspace{0pt} shows more absorption than the other two lines. In fact, to some extent this is seen also for the other two stars. 

There are also C\textsubscript{2} Phillips system lines visible in the observed spectra. These also have variable emission/absorption components; however, their profiles are often blended with narrow blueshifted absorptions that are formed in the AGB ejecta. Some C\textsubscript{2} Phillips lines blend the CO (3,0) band head.

\section{Discussion}
\label{sec:Discussion}

The CO $\Delta v=3$ lines in the three observed post-AGB stars show a tendency to have the strongest emission when a star is close to pulsation phase light maximum and the weakest emission or strongest absorption in light minimum phase. Also, shapes, widths, and positions of CO lines change during the pulsation. The CO features, when they are purely in emission or absorption, have FWHM in the range from around 14 to 17 km/s and they are positioned at around systemic velocity or blueshifted by few km/s. Generally, for CN Red system and CO lines the velocities and widths approximately match. However, CN Red system lines tend to have stronger absorption. In some spectra, molecular lines are P Cygni or inverse P Cygni-like showing both emission and absorption no farther than roughly 10 km/s from the systemic velocity. For IRAS 22272+5435 and IRAS Z02229+6208 at phases around $\varphi$=0.7-0.8, CN lines are of P Cygni-like shape while CO lines are purely in emission. While there is a similarity in molecular line shapes and positions between 2022 and 2025 spectra of IRAS Z02229+6208 that correspond to similar pulsation phases, it is likely that the variation of these spectral features does not strictly repeat at different pulsation cycles. \citet{Pukitis2023} reported some spectra of IRAS 22272+5435 with molecular lines in absorption despite corresponding to pulsation phases close light maximum. Widths of the variable molecular features suggest that they are neither of circumstellar nor photospheric origin. For the former, FWHM is around 5 km/s or lower. Photospheric absorptions have FWHM of 20 km/s or higher. While the absorption intensities at the very beginning of C\textsubscript{2} and CO band heads are similar in synthetic and observed spectra at minimum light phase of IRAS 22272+5435, lines that are farther form the start of the band heads are significantly weaker in the synthetic spectrum. Intensities of CN lines are not matched even at the start of the band heads. Also, positions of molecular features do not match the photospheric velocity, in general. Thus, the site of formation seems to be above the photosphere, in the extended atmosphere. The reason why CO line variability is similar to but does not match changes in CN lines is most likely due to different heights of formation caused by different excitation energies. This could also be the reason why there is a difference in variability for different CN lines (Figure \ref{fig:CNindivid} and \citet{Zacs2023}) and between CO $\Delta v=3$ and $\Delta v=2$ features \citep{Raman2008}. It has been suggested that CO emission in post-AGB objects might be related to a circumstellar disc \citep{Gledhill2011}. However, this is unlikely the case for the three observed stars as they have a shell-type spectral energy distribution suggesting the lack of such disc \citep{Gezer2015}.

\citet{Oudmaijer1995} observed CO first overtone line emission near light maximum and absorption near light minimum for the post-AGB star AC Herculis\footnote{It is an RV Tauri type variable - its pulsation nature is different than that of the three stars analysed here.}. The emission was seen when the photopshere was contracting and absorption - during expansion. They proposed the idea that CO emission is formed in a shell that detaches from the photosphere after the expansion. The observed molecular line variability is in agreement with this interpretation in the case of IRAS 2272+5435. However, emission in both CO $\Delta v=3$ and CN lines is seen in 2025 October spectra of IRAS Z02229+6208 when the photosphere was expanding. Also, positions of molecular absorption lines in 2022 \textcolor{blue}{July 22} spectrum of IRAS 20000+3239 show expansion during contraction of the photosphere suggesting that the region where molecular absorption forms is detached. This shows that the dynamics in the extended atmospheres of post-AGB stars is more complex than in the \citet{Oudmaijer1995} scenario and detailed modelling of this region is necessary to interpret the observed molecular line variability. Such modelling has not been done for post-AGB stars. However, formation of the variable C\textsubscript{2} and CN molecular features in IRAS 22272+5435 was discussed by \citet{Pukitis2023} in the context of dynamic AGB star models. In short, the variability could be caused by both upward and downward moving matter in the extended atmosphere that is deviated from spherical symmetry due to convection. Indeed, the observed positions of molecular absorption and P Cygni-type profiles of these lines point to outward motion of matter. The observed inverse P Cygni-type profiles in molecular lines suggest that matter is falling onto the star. However, the explanation for the apparent connection of molecular variability with the pulsation of the star is still unclear. Appearance of CO second overtone emission at phases near maximum light has been observed in the yellow hypergiant $\rho$ Cassiopeiae, and it has been explained by pulsation-driven shock waves in atmospheric layers above the photosphere \citep{Gorlova2006}. \citet{Mozurkevizh1987} linked the variable CO emission in the RV Tauri-type star R Scuti with a shock wave that travels trough the extend atmosphere once per pulsation cycle. Such a shock could be responsible for molecular excitation in the three post-AGB stars analysed here.

\begin{acknowledgments}
The research is financed by the Recovery and Resilience Facility project "Internal and External Consolidation of the University of Latvia" (No.5.2.1.1.i.0/2/24/I/CFLA/007).

This research received funding from the European Union's Horizon 2020 research and innovation program under grant agreement No. 101004719 (OPTICON-Radionet Pilot). 

Based on observations collected at Centro Astronómico Hispano en Andalucía (CAHA) at Calar Alto, proposals 22B-3.5-055 and 25B-3.5-014, operated jointly by Junta de Andalucía and Consejo Superior de Investigaciones Científicas (IAA-CSIC).

Based on data from the CAHA Archive at CAB (INTA-CSIC). The CAHA Archive is part of the Spanish Virtual Observatory project funded by MCIN/AEI/10.13039/501100011033 through grant PID2020-112949GB-I00.

This work has made use of the VALD database, operated at Uppsala University, the Institute of Astronomy RAS in Moscow, and the University of Vienna.
\end{acknowledgments}

\software{DECH \citep{Galazutdinov2022}, SPECTRUM \citep{Gray1994}}

\bibliography{sample631}{}

@ARTICLE{Kochanek2017,
       author = {{Kochanek}, C.~S. and {Shappee}, B.~J. and {Stanek}, K.~Z. and {Holoien}, T.~W. -S. and {Thompson}, Todd A. and {Prieto}, J.~L. and {Dong}, Subo and {Shields}, J.~V. and {Will}, D. and {Britt}, C. and {Perzanowski}, D. and {Pojma{\'n}ski}, G.},
        title = "{The All-Sky Automated Survey for Supernovae (ASAS-SN) Light Curve Server v1.0}",
      journal = {\pasp},
         year = 2017,
        month = oct,
       volume = {129},
       number = {980},
        pages = {104502},
          doi = {10.1088/1538-3873/aa80d9},
archivePrefix = {arXiv},
       eprint = {1706.07060},
 primaryClass = {astro-ph.SR},
       adsurl = {https://ui.adsabs.harvard.edu/abs/2017PASP..129j4502K}
}

@ARTICLE{Pukitis2023,
       author = {{Pu{\c{k}}{\={\i}}tis}, K{\={a}}rlis and {Za{\v{c}}s}, Laimons and {Sperauskas}, Julius},
        title = "{Episodes of Molecular Emission in the Optical Spectrum of IRAS 22272+5435}",
      journal = {\apj},
         year = 2023,
        month = may,
       volume = {948},
       number = {1},
          eid = {70},
        pages = {70},
          doi = {10.3847/1538-4357/acc52b},
       adsurl = {https://ui.adsabs.harvard.edu/abs/2023ApJ...948...70P}
}

@ARTICLE{Klochkova2006,
       author = {{Klochkova}, Valentina G. and {Kipper}, T{\~o}nu},
        title = "{Optical Spectroscopy of the Post-AGB Carbon Star Cgcs 6857 = IRAS 20000+3239}",
      journal = {Baltic Astronomy},
         year = 2006,
        month = jan,
       volume = {15},
        pages = {395-404},
       adsurl = {https://ui.adsabs.harvard.edu/abs/2006BaltA..15..395K}
}

@ARTICLE{Zacs2025,
       author = {{Za{\v{c}}s}, Laimons and {Pu{\c{k}}{\={\i}}tis}, K{\={a}}rlis},
        title = "{Variability of physical parameters of IRAS 22272<inline-formula id=``IEq1''><mml:math id=``IEq1\_Math''><mml:mo>+</mml:mo></mml:math></inline-formula>5435 during the pulsation cycle}",
      journal = {Journal of Astrophysics and Astronomy},
         year = 2025,
        month = jan,
       volume = {46},
       number = {1},
          eid = {11},
        pages = {11},
          doi = {10.1007/s12036-024-10037-5},
       adsurl = {https://ui.adsabs.harvard.edu/abs/2025JApA...46...11Z}
}

@ARTICLE{Oudmaijer1995,
       author = {{Oudmaijer}, R.~D. and {Waters}, L.~B.~F.~M. and {van der Veen}, W.~E.~C.~J. and {Geballe}, T.~R.},
        title = "{Near-infrared spectroscopy of post-AGB stars.}",
      journal = {\aap},
         year = 1995,
        month = jul,
       volume = {299},
        pages = {69},
       adsurl = {https://ui.adsabs.harvard.edu/abs/1995A&A...299...69O}
}

@ARTICLE{Mozurkevizh1987,
       author = {{Mozurkewich}, David and {Gehrz}, R.~D. and {Hinkle}, K.~H. and {Lambert}, D.~L.},
        title = "{Velocity Structure of Stellar Atmospheres: R Scuti}",
      journal = {\apj},
         year = 1987,
        month = mar,
       volume = {314},
        pages = {242},
          doi = {10.1086/165053},
       adsurl = {https://ui.adsabs.harvard.edu/abs/1987ApJ...314..242M}
}

@ARTICLE{Hrivnak1994,
       author = {{Hrivnak}, Bruce J. and {Kwok}, Sun and {Geballe}, T.~R.},
        title = "{Near-Infrared Spectroscopy of Proto--Planetary Nebulae}",
      journal = {\apj},
         year = 1994,
        month = jan,
       volume = {420},
        pages = {783},
          doi = {10.1086/173602},
       adsurl = {https://ui.adsabs.harvard.edu/abs/1994ApJ...420..783H}
}

@ARTICLE{Gorlova2006,
       author = {{Gorlova}, Nadya and {Lobel}, Alex and {Burgasser}, Adam J. and {Rieke}, George H. and {Ilyin}, Ilya and {Stauffer}, John R.},
        title = "{On the CO Near-Infrared Band and the Line-splitting Phenomenon in the Yellow Hypergiant {\ensuremath{\rho}} Cassiopeiae}",
      journal = {\apj},
         year = 2006,
        month = nov,
       volume = {651},
       number = {2},
        pages = {1130-1150},
          doi = {10.1086/507590},
archivePrefix = {arXiv},
       eprint = {astro-ph/0607158},
 primaryClass = {astro-ph},
       adsurl = {https://ui.adsabs.harvard.edu/abs/2006ApJ...651.1130G}
}

@ARTICLE{Raman2008,
       author = {{Venkata Raman}, V. and {Anandarao}, B.~G.},
        title = "{Infrared spectroscopic study of a selection of AGB and post-AGB stars}",
      journal = {\mnras},
         year = 2008,
        month = apr,
       volume = {385},
       number = {2},
        pages = {1076-1086},
          doi = {10.1111/j.1365-2966.2008.12915.x},
archivePrefix = {arXiv},
       eprint = {0801.0910},
 primaryClass = {astro-ph},
       adsurl = {https://ui.adsabs.harvard.edu/abs/2008MNRAS.385.1076V}
}

@ARTICLE{Gledhill2011,
       author = {{Gledhill}, T.~M. and {Forde}, K.~P. and {Lowe}, K.~T.~E. and {Smith}, M.~D.},
        title = "{Integral field spectroscopy of H$_{2}$ and CO emission in IRAS 18276-1431: evidence for ongoing post-AGB mass-loss}",
      journal = {\mnras},
         year = 2011,
        month = mar,
       volume = {411},
       number = {3},
        pages = {1453-1466},
          doi = {10.1111/j.1365-2966.2010.17779.x},
archivePrefix = {arXiv},
       eprint = {1009.5608},
 primaryClass = {astro-ph.SR},
       adsurl = {https://ui.adsabs.harvard.edu/abs/2011MNRAS.411.1453G}
}

@ARTICLE{Bertolami2016,
       author = {{Miller Bertolami}, Marcelo Miguel},
        title = "{New models for the evolution of post-asymptotic giant branch stars and central stars of planetary nebulae}",
      journal = {\aap},
         year = 2016,
        month = apr,
       volume = {588},
          eid = {A25},
        pages = {A25},
          doi = {10.1051/0004-6361/201526577},
archivePrefix = {arXiv},
       eprint = {1410.1679},
 primaryClass = {astro-ph.SR},
       adsurl = {https://ui.adsabs.harvard.edu/abs/2016A&A...588A..25M}
}

@ARTICLE{Zacs2023,
       author = {{Za{\v{c}}s}, Laimons and {Pu{\c{k}}{\={\i}}tis}, K{\={a}}rlis},
        title = "{Pulsation-induced Spectroscopic Variability of IRAS Z02229+6208}",
      journal = {\apj},
         year = 2023,
        month = jul,
       volume = {952},
       number = {1},
          eid = {49},
        pages = {49},
          doi = {10.3847/1538-4357/acdcfe},
       adsurl = {https://ui.adsabs.harvard.edu/abs/2023ApJ...952...49Z}
}

@ARTICLE{Reddy1999,
       author = {{Reddy}, Bacham E. and {Bakker}, Eric J. and {Hrivnak}, Bruce J.},
        title = "{An Abundance Analysis of Two Carbon-rich Proto-Planetary Nebulae: IRAS Z02229+6208 and IRAS 07430+1115}",
      journal = {\apj},
         year = 1999,
        month = oct,
       volume = {524},
       number = {2},
        pages = {831-848},
          doi = {10.1086/307858},
       adsurl = {https://ui.adsabs.harvard.edu/abs/1999ApJ...524..831R}
}

@ARTICLE{Hrivnak2022,
       author = {{Hrivnak}, Bruce J. and {Lu}, Wenxian and {Bakke}, William C. and {Grimm}, Peyton J.},
        title = "{Variability in Protoplanetary Nebulae. IX. Evidence for Evolution in a Decade}",
      journal = {\apj},
         year = 2022,
        month = nov,
       volume = {939},
       number = {1},
          eid = {32},
        pages = {32},
          doi = {10.3847/1538-4357/ac938a},
archivePrefix = {arXiv},
       eprint = {2210.00103},
 primaryClass = {astro-ph.SR},
       adsurl = {https://ui.adsabs.harvard.edu/abs/2022ApJ...939...32H}
}

@INPROCEEDINGS{Quirrenbach2014,
       author = {{Quirrenbach}, A. and {Amado}, P.~J. and {Caballero}, J.~A. and {Mundt}, R. and {Reiners}, A. and {Ribas}, I. and {Seifert}, W. and {Abril}, M. and {Aceituno}, J. and {Alonso-Floriano}, F.~J. and {Ammler-von Eiff}, M. and {Antona Jim{\'e}nez}, R. and {Anwand-Heerwart}, H. and {Azzaro}, M. and {Bauer}, F. and {Barrado}, D. and {Becerril}, S. and {B{\'e}jar}, V.~J.~S. and {Ben{\'\i}tez}, D. and {Berdi{\~n}as}, Z.~M. and {C{\'a}rdenas}, M.~C. and {Casal}, E. and {Claret}, A. and {Colom{\'e}}, J. and {Cort{\'e}s-Contreras}, M. and {Czesla}, S. and {Doellinger}, M. and {Dreizler}, S. and {Feiz}, C. and {Fern{\'a}ndez}, M. and {Galad{\'\i}}, D. and {G{\'a}lvez-Ortiz}, M.~C. and {Garc{\'\i}a-Piquer}, A. and {Garc{\'\i}a-Vargas}, M.~L. and {Garrido}, R. and {Gesa}, L. and {G{\'o}mez Galera}, V. and {Gonz{\'a}lez {\'A}lvarez}, E. and {Gonz{\'a}lez Hern{\'a}ndez}, J.~I. and {Gr{\"o}zinger}, U. and {Gu{\`a}rdia}, J. and {Guenther}, E.~W. and {de Guindos}, E. and {Guti{\'e}rrez-Soto}, J. and {Hagen}, H. -J. and {Hatzes}, A.~P. and {Hauschildt}, P.~H. and {Helmling}, J. and {Henning}, T. and {Hermann}, D. and {Hern{\'a}ndez Casta{\~n}o}, L. and {Herrero}, E. and {Hidalgo}, D. and {Holgado}, G. and {Huber}, A. and {Huber}, K.~F. and {Jeffers}, S. and {Joergens}, V. and {de Juan}, E. and {Kehr}, M. and {Klein}, R. and {K{\"u}rster}, M. and {Lamert}, A. and {Lalitha}, S. and {Laun}, W. and {Lemke}, U. and {Lenzen}, R. and {L{\'o}pez del Fresno}, Mauro and {L{\'o}pez Mart{\'\i}}, B. and {L{\'o}pez-Santiago}, J. and {Mall}, U. and {Mandel}, H. and {Mart{\'\i}n}, E.~L. and {Mart{\'\i}n-Ruiz}, S. and {Mart{\'\i}nez-Rodr{\'\i}guez}, H. and {Marvin}, C.~J. and {Mathar}, R.~J. and {Mirabet}, E. and {Montes}, D. and {Morales Mu{\~n}oz}, R. and {Moya}, A. and {Naranjo}, V. and {Ofir}, A. and {Oreiro}, R. and {Pall{\'e}}, E. and {Panduro}, J. and {Passegger}, V. -M. and {P{\'e}rez-Calpena}, A. and {P{\'e}rez Medialdea}, D. and {Perger}, M. and {Pluto}, M. and {Ram{\'o}n}, A. and {Rebolo}, R. and {Redondo}, P. and {Reffert}, S. and {Reinhardt}, S. and {Rhode}, P. and {Rix}, H. -W. and {Rodler}, F. and {Rodr{\'\i}guez}, E. and {Rodr{\'\i}guez-L{\'o}pez}, C. and {Rodr{\'\i}guez-P{\'e}rez}, E. and {Rohloff}, R. -R. and {Rosich}, A. and {S{\'a}nchez-Blanco}, E. and {S{\'a}nchez Carrasco}, M.~A. and {Sanz-Forcada}, J. and {Sarmiento}, L.~F. and {Sch{\"a}fer}, S. and {Schiller}, J. and {Schmidt}, C. and {Schmitt}, J.~H.~M.~M. and {Solano}, E. and {Stahl}, O. and {Storz}, C. and {St{\"u}rmer}, J. and {Su{\'a}rez}, J.~C. and {Ulbrich}, R.~G. and {Veredas}, G. and {Wagner}, K. and {Winkler}, J. and {Zapatero Osorio}, M.~R. and {Zechmeister}, M. and {Abell{\'a}n de Paco}, F.~J. and {Anglada-Escud{\'e}}, G. and {del Burgo}, C. and {Klutsch}, A. and {Lizon}, J.~L. and {L{\'o}pez-Morales}, M. and {Morales}, J.~C. and {Perryman}, M.~A.~C. and {Tulloch}, S.~M. and {Xu}, W.},
        title = "{CARMENES instrument overview}",
    booktitle = {Ground-based and Airborne Instrumentation for Astronomy V},
         year = 2014,
       editor = {{Ramsay}, Suzanne K. and {McLean}, Ian S. and {Takami}, Hideki},
       series = {Society of Photo-Optical Instrumentation Engineers (SPIE) Conference Series},
       volume = {9147},
        month = jul,
          eid = {91471F},
        pages = {91471F},
          doi = {10.1117/12.2056453},
       adsurl = {https://ui.adsabs.harvard.edu/abs/2014SPIE.9147E..1FQ}
}

@ARTICLE{Caballero2025,
       author = {{Caballero}, Jos{\'e} A. and {Seifert}, Walter and {Quirrenbach}, Andreas and {Amado}, Pedro J. and {Ribas}, Ignasi and {Reiners}, Ansgar},
        title = "{CARMENES as an Instrument for Exoplanet Research}",
      journal = {arXiv e-prints},
         year = 2025,
        month = mar,
          eid = {arXiv:2503.05501},
        pages = {arXiv:2503.05501},
          doi = {10.48550/arXiv.2503.05501},
archivePrefix = {arXiv},
       eprint = {2503.05501},
 primaryClass = {astro-ph.EP},
       adsurl = {https://ui.adsabs.harvard.edu/abs/2025arXiv250305501C}
}

@INPROCEEDINGS{Caballero2016,
       author = {{Caballero}, J.~A. and {Gu{\`a}rdia}, J. and {L{\'o}pez del Fresno}, M. and {Zechmeister}, M. and {de Juan}, E. and {Alonso-Floriano}, F.~J. and {Amado}, P.~J. and {Colom{\'e}}, J. and {Cort{\'e}s-Contreras}, M. and {Garc{\'\i}a-Piquer}, {\'A}. and {Gesa}, L. and {de Guindos}, E. and {Hagen}, H. -J. and {Helmling}, J. and {Hern{\'a}ndez Casta{\~n}o}, L. and {K{\"u}rster}, M. and {L{\'o}pez-Santiago}, J. and {Montes}, D. and {Morales Mu{\~n}oz}, R. and {Pavlov}, A. and {Quirrenbach}, A. and {Reiners}, A. and {Ribas}, I. and {Seifert}, W. and {Solano}, E.},
        title = "{CARMENES: data flow}",
    booktitle = {Observatory Operations: Strategies, Processes, and Systems VI},
         year = 2016,
       editor = {{Peck}, Alison B. and {Seaman}, Robert L. and {Benn}, Chris R.},
       series = {Society of Photo-Optical Instrumentation Engineers (SPIE) Conference Series},
       volume = {9910},
        month = jul,
          eid = {99100E},
        pages = {99100E},
          doi = {10.1117/12.2233574},
       adsurl = {https://ui.adsabs.harvard.edu/abs/2016SPIE.9910E..0EC}
}

@ARTICLE{Shappee2014,
       author = {{Shappee}, B.~J. and {Prieto}, J.~L. and {Grupe}, D. and {Kochanek}, C.~S. and {Stanek}, K.~Z. and {De Rosa}, G. and {Mathur}, S. and {Zu}, Y. and {Peterson}, B.~M. and {Pogge}, R.~W. and {Komossa}, S. and {Im}, M. and {Jencson}, J. and {Holoien}, T.~W. -S. and {Basu}, U. and {Beacom}, J.~F. and {Szczygie{\l}}, D.~M. and {Brimacombe}, J. and {Adams}, S. and {Campillay}, A. and {Choi}, C. and {Contreras}, C. and {Dietrich}, M. and {Dubberley}, M. and {Elphick}, M. and {Foale}, S. and {Giustini}, M. and {Gonzalez}, C. and {Hawkins}, E. and {Howell}, D.~A. and {Hsiao}, E.~Y. and {Koss}, M. and {Leighly}, K.~M. and {Morrell}, N. and {Mudd}, D. and {Mullins}, D. and {Nugent}, J.~M. and {Parrent}, J. and {Phillips}, M.~M. and {Pojmanski}, G. and {Rosing}, W. and {Ross}, R. and {Sand}, D. and {Terndrup}, D.~M. and {Valenti}, S. and {Walker}, Z. and {Yoon}, Y.},
        title = "{The Man behind the Curtain: X-Rays Drive the UV through NIR Variability in the 2013 Active Galactic Nucleus Outburst in NGC 2617}",
      journal = {\apj},
         year = 2014,
        month = jun,
       volume = {788},
       number = {1},
          eid = {48},
        pages = {48},
          doi = {10.1088/0004-637X/788/1/48},
archivePrefix = {arXiv},
       eprint = {1310.2241},
 primaryClass = {astro-ph.HE},
       adsurl = {https://ui.adsabs.harvard.edu/abs/2014ApJ...788...48S}
}

@ARTICLE{Galazutdinov2022,
       author = {{Galazutdinov}, G.~A.},
        title = "{DECH: A Software Package for Astronomical Spectral Data Processing and Analysis}",
      journal = {Astrophysical Bulletin},
         year = 2022,
        month = dec,
       volume = {77},
       number = {4},
        pages = {519-529},
          doi = {10.1134/S1990341322040034},
       adsurl = {https://ui.adsabs.harvard.edu/abs/2022AstBu..77..519G}
}

@ARTICLE{Hrivnak2013,
       author = {{Hrivnak}, Bruce J. and {Lu}, Wenxian and {Sperauskas}, Julius and {Van Winckel}, Hans and {Bohlender}, David and {Za{\v{c}}s}, Laimons},
        title = "{Studies of Variability in Proto-planetary Nebulae. II. Light and Velocity Curve Analyses of IRAS 22272+5435 and 22223+4327}",
      journal = {\apj},
         year = 2013,
        month = apr,
       volume = {766},
       number = {2},
          eid = {116},
        pages = {116},
          doi = {10.1088/0004-637X/766/2/116},
archivePrefix = {arXiv},
       eprint = {1310.8558},
 primaryClass = {astro-ph.SR},
       adsurl = {https://ui.adsabs.harvard.edu/abs/2013ApJ...766..116H}
}

@ARTICLE{Hrivnak1999,
       author = {{Hrivnak}, Bruce J. and {Kwok}, Sun},
        title = "{Discovery of Two New, Carbon-rich Proto-Planetary Nebulae:IRAS Z02229+6208 and IRAS 07430+1115}",
      journal = {\apj},
         year = 1999,
        month = mar,
       volume = {513},
       number = {2},
        pages = {869-878},
          doi = {10.1086/306873},
       adsurl = {https://ui.adsabs.harvard.edu/abs/1999ApJ...513..869H}
}

@ARTICLE{Hrivnak2005,
       author = {{Hrivnak}, Bruce J. and {Bieging}, John H.},
        title = "{CO J = 2-1 and 4-3 Observations of Proto-planetary Nebulae: Time-variable Mass Loss}",
      journal = {\apj},
         year = 2005,
        month = may,
       volume = {624},
       number = {1},
        pages = {331-351},
          doi = {10.1086/428894},
       adsurl = {https://ui.adsabs.harvard.edu/abs/2005ApJ...624..331H}
}

@ARTICLE{Contreras2012,
       author = {{S{\'a}nchez Contreras}, C. and {Sahai}, R.},
        title = "{OPACOS: OVRO Post-AGB CO (1-0) Emission Survey. I. Data and Derived Nebular Parameters}",
      journal = {\apjs},
         year = 2012,
        month = nov,
       volume = {203},
       number = {1},
          eid = {16},
        pages = {16},
          doi = {10.1088/0067-0049/203/1/16},
       adsurl = {https://ui.adsabs.harvard.edu/abs/2012ApJS..203...16S}
}

@ARTICLE{Li2015,
       author = {{Li}, Gang and {Gordon}, Iouli E. and {Rothman}, Laurence S. and {Tan}, Yan and {Hu}, Shui-Ming and {Kassi}, Samir and {Campargue}, Alain and {Medvedev}, Emile S.},
        title = "{Rovibrational Line Lists for Nine Isotopologues of the CO Molecule in the X $^{1}${\ensuremath{\Sigma}}$^{+}$ Ground Electronic State}",
      journal = {\apjs},
         year = 2015,
        month = jan,
       volume = {216},
       number = {1},
          eid = {15},
        pages = {15},
          doi = {10.1088/0067-0049/216/1/15},
       adsurl = {https://ui.adsabs.harvard.edu/abs/2015ApJS..216...15L}
}

@ARTICLE{Piskunov1995,
       author = {{Piskunov}, N.~E. and {Kupka}, F. and {Ryabchikova}, T.~A. and {Weiss}, W.~W. and {Jeffery}, C.~S.},
        title = "{VALD: The Vienna Atomic Line Data Base.}",
      journal = {\aaps},
         year = 1995,
        month = sep,
       volume = {112},
        pages = {525},
       adsurl = {https://ui.adsabs.harvard.edu/abs/1995A&AS..112..525P}
}

@ARTICLE{Kupka1999,
       author = {{Kupka}, F. and {Piskunov}, N. and {Ryabchikova}, T.~A. and {Stempels}, H.~C. and {Weiss}, W.~W.},
        title = "{VALD-2: Progress of the Vienna Atomic Line Data Base}",
      journal = {\aaps},
         year = 1999,
        month = jul,
       volume = {138},
        pages = {119-133},
          doi = {10.1051/aas:1999267},
       adsurl = {https://ui.adsabs.harvard.edu/abs/1999A&AS..138..119K}
}

@ARTICLE{Zacs2016,
       author = {{Za{\v{c}}s}, Laimons and {Musaev}, Faig and {Kaminsky}, Bogdan and {Pavlenko}, Yakiv and {Grankina}, Aija and {Sperauskas}, Julius and {Hrivnak}, Bruce J.},
        title = "{Spectroscopic Variability of IRAS 22272+5435}",
      journal = {\apj},
         year = 2016,
        month = jan,
       volume = {816},
       number = {1},
          eid = {3},
        pages = {3},
          doi = {10.3847/0004-637X/816/1/3},
archivePrefix = {arXiv},
       eprint = {1511.03450},
 primaryClass = {astro-ph.SR},
       adsurl = {https://ui.adsabs.harvard.edu/abs/2016ApJ...816....3Z}
}

@ARTICLE{DeSmedt2016,
       author = {{De Smedt}, K. and {Van Winckel}, H. and {Kamath}, D. and {Siess}, L. and {Goriely}, S. and {Karakas}, A.~I. and {Manick}, R.},
        title = "{Detailed homogeneous abundance studies of 14 Galactic s-process enriched post-AGB stars: In search of lead (Pb)}",
      journal = {\aap},
         year = 2016,
        month = mar,
       volume = {587},
          eid = {A6},
        pages = {A6},
          doi = {10.1051/0004-6361/201527430},
archivePrefix = {arXiv},
       eprint = {1512.05393},
 primaryClass = {astro-ph.SR},
       adsurl = {https://ui.adsabs.harvard.edu/abs/2016A&A...587A...6D}
}

@ARTICLE{Gray1994,
       author = {{Gray}, R.~O. and {Corbally}, C.~J.},
        title = "{The Calibration of MK Spectral Classes Using Spectral Synthesis. I. The Effective Temperature Calibration of Dwarf Stars}",
      journal = {\aj},
         year = 1994,
        month = feb,
       volume = {107},
        pages = {742},
          doi = {10.1086/116893},
       adsurl = {https://ui.adsabs.harvard.edu/abs/1994AJ....107..742G}
}

@ARTICLE{Gezer2015,
       author = {{Gezer}, I. and {Van Winckel}, H. and {Bozkurt}, Z. and {De Smedt}, K. and {Kamath}, D. and {Hillen}, M. and {Manick}, R.},
        title = "{The WISE view of RV Tauri stars}",
      journal = {\mnras},
         year = 2015,
        month = oct,
       volume = {453},
       number = {1},
        pages = {133-146},
          doi = {10.1093/mnras/stv1627},
archivePrefix = {arXiv},
       eprint = {1507.04175},
 primaryClass = {astro-ph.SR},
       adsurl = {https://ui.adsabs.harvard.edu/abs/2015MNRAS.453..133G}
}
\bibliographystyle{aasjournal}

\end{document}